%% file: GOSPA_Fusion31Jan2017.tex
\documentclass[journal]{IEEEtran}
\usepackage{amsthm, amssymb, amsfonts,algorithm, algpseudocode, array}
\usepackage[cmex10]{amsmath}
\usepackage[english]{babel}
\usepackage{blindtext}
\usepackage{cancel,cite,color,comment}
\usepackage[margin=8pt,font=footnotesize]{caption}
\usepackage{enumerate,esint}
\usepackage{float}
\usepackage[T1]{fontenc}
\usepackage{graphicx}
\usepackage[latin9]{inputenc}
\usepackage{multirow,multicol}
\usepackage{slashbox,soul}
\usepackage{tikz,tgtermes}
\usepackage[labelformat=simple]{subcaption}
\usepackage{epstopdf}
\usepackage{pgfplots}

\usetikzlibrary{shapes.multipart}
\allowdisplaybreaks
\newcommand{\dG}{{d_{p}^{(c, \alpha)}}}
\newcommand{\dGpone}{{d_{p}^{(c,1)}}}

\newcommand{\dGtwo}{{d_{1}^{(c,2)}}}
\newcommand{\dGtwotwo}{{d_{2}^{(c,2)}}}
\newcommand{\dGptwo}{{d_{p}^{(c,2)}}}

\newcommand{\dc}{{d^{(c)}}}
\newcommand{\nn}{{\nonumber}}
\newcommand{\RN}{\mathbb{R}^N}
\newcommand{\E}{\mathbb{E}}
\newcommand{\nX}{{|X|}}
\newcommand{\nY}{{|Y|}}
\newcommand{\nZ}{{|Z|}}
\newcommand{\eye}{{I}}

\renewcommand{\qed}{{\hfill$\square$}}

\theoremstyle{plain}
\newtheorem{thm}{Theorem}

\newtheorem{prop}[thm]{Proposition}

\theoremstyle{definition}
\newtheorem{defn}{Definition}

\newtheorem{example}{Example}
\theoremstyle{remark}

\begin{document}
\title{Generalized optimal sub-pattern assignment metric}
\author{Abu Sajana Rahmathullah, Ángel F. García-Fernández, Lennart
	Svensson \thanks{A.S. Rahmathullah is with Zenuity AB, Gothenburg, Sweden (email: abusajana@gmail.com). Á. F. García-Fernández is with the Department of Electrical Engineering and Automation, Aalto University, 02150 Espoo, Finland (email: angel.garciafernandez@aalto.fi). L. Svensson is with the Department of Signals and Systems, Chalmers University of Technology, SE-412 96 Gothenburg, Sweden (email: lennart.svensson@chalmers.se).}}
\date{\today}

\maketitle
\begin{abstract}
This paper presents the generalized optimal sub-pattern assignment (GOSPA) metric on the space of finite sets of targets. Compared to the well-established optimal sub-pattern assignment  (OSPA) metric, GOSPA is not normalised by the cardinality of the largest set and it penalizes cardinality errors differently, which enables us to express it as an optimisation over assignments instead of permutations. An important consequence of this is that GOSPA allows us to penalize localization errors for detected targets and the errors due to missed and false targets, as indicated by traditional multiple target tracking (MTT) performance measures, in a sound manner. In addition, we extend the GOSPA metric to the space of random finite sets, which is important to evaluate MTT algorithms via simulations in a rigorous way.
\end{abstract}

\begin{IEEEkeywords}
Multiple target tracking, metric, random finite sets, optimal sub-pattern assignment metric.
\end{IEEEkeywords}
\section{Introduction}
Multiple target tracking (MTT) algorithms sequentially estimate a set of targets, which  appear, move and disappear from a scene, given noisy sensor observations \cite{mahler2007statistical}. In order to assess and compare the performance of MTT algorithms, one needs to compute the similarity between the ground truth and the estimated set. Traditionally, MTT performance assessment has been based on intuitive concepts such as localization error for properly detected targets and costs for missed targets and false targets \cite[Sec. 13.6]{blackman1999design}, \cite{fridling1991performance,rothrock2000performance,mabbs1993performance,Drummond92}. These concepts are appealing and practical for radar operators but the way they have been quantified to measure error has been ad-hoc. 
%This is not theoretically satisfying and could therefore lead to unexpected results in some situations. 

With the advent of the random finite set (RFS) framework for MTT \cite{mahler2007statistical}, it has been possible to design and define the errors in a mathematically sound way, without ad-hoc mechanisms.  In this framework, at any given time, the ground truth is a set that contains the true target states and the estimate is a set that contains the estimated target states. The error is then the distance  between these two sets according to a metric, which satisfies the properties of non-negativity, definiteness, symmetry and triangle inequality  \cite[Sec.~ 1]{ristic2010performance}, \cite[Sec.~6.2.1]{mahler2014advances}. %Specifically, the need for the triangle inequality is well motivated in: if there is an estimate $A$ which is close to the ground truth and another estimate $B$ which is close to $A$, then the triangle inequality ensures that the estimate $B$ is also close to the ground truth, which agrees with the intuition.

The Hausdorff metric  \cite{hoffman2004multitarget} and the Wasserstein metric  \cite{hoffman2004multitarget}  (also referred to as optimal mass transfer metric in \cite{schuhmacher2008consistent}) were the first metrics on the space of finite sets of targets for MTT. However, the former has been shown to be insensitive to cardinality mismatches and the latter lacks a consistent physical interpretation when the states have different cardinalities \cite{schuhmacher2008consistent}. In \cite{schuhmacher2008consistent}, the optimal sub-pattern assignment (OSPA) metric was proposed to address these issues. OSPA optimally assigns all targets in the smallest set to targets in the other set and computes a localization cost based on this assignment. The rest of the targets are accounted for by a cardinality mismatch penalty. The OSPA metric has also been adapted to handle sets of labelled targets \cite{ristic2011metric}. 

%This metric is also easily computed and has been used widely in the tracking community for performance evaluation and to obtain estimators \cite{garcia2014bayesian, williams2015efficient, baum2015wasserstein}. 

We argue that it is more desirable to have a metric that accounts for the costs mentioned in traditional MTT performance assessment methods (localization error for properly detected targets and false and missed targets) rather than localization error for the targets in the smallest set and cardinality mismatch,  which is a mathematical concept that is more related to the RFS formulation of MTT problem than the original MTT problem itself. For example, OSPA does not encourage trackers to have as few false and missed targets as possible. 

In this paper, we propose such a metric: the generalized OSPA (GOSPA) metric, which is able to penalize localization errors for properly detected targets, missed targets and false targets. In order to obtain this metric, we first generalize the unnormalized OSPA by including an additional parameter that enables us to select the cardinality mismatch cost from a range of values. Then, we show that for a specific selection of this parameter, the GOSPA metric is a sum of localization errors for the properly detected targets and a penalty for missed and false targets, as in traditional MTT performance assessment algorithms. Importantly, this implies that we now have a metric that satisfies the fundamental properties of metrics and the intuitive, classical notions of how MTT algorithms should be evaluated \cite[13.6]{blackman1999design}. After we derived the GOSPA metric, it has been used in a separate performance evaluation \cite{xia2017performance}, which illustrates the usefulness of GOSPA for analysing how the number of missed and false targets contribute to the total loss. 

We also extend the metric to random sets of targets. This extension has received less attention in the MTT literature despite its significance for performance evaluation. All the above-mentioned metrics assume that the ground truth and the estimates are known. However, in the RFS framework, the ground truth is not known but is modelled as a random finite set \cite{mahler2007statistical}. Also, algorithm evaluation is usually performed by averaging the error of the estimates for different measurements obtained by Monte Carlo simulation. This implies that the estimates are also RFSs, so it is important to have a metric that considers RFSs rather than finite sets. In the literature, there is no formal treatment of this problem to our knowledge. In this paper, we fill this gap by showing that the mean GOSPA and root mean square GOSPA are metrics for RFSs of targets. 

The outline of the rest of the paper is as follows. In Section~\ref{sec:GOSPA}, we present the GOSPA and its most appropriate form for MTT. In Section~\ref{sec:perf}, we extend it to RFS of targets. In Section~\ref{sec:ill}, we illustrate that the proposed choice of GOSPA provides expected  results compared to OSPA and unnormalized OSPA. Finally, conclusions are drawn in Section~\ref{sec:concl}.
\section{Generalized OSPA metric}\label{sec:GOSPA}
In this section, we present the generalized OSPA (GOSPA) metric to measure the distance between finite sets of targets. %We also compare GOSPA metric with OSPA and discuss the drawback of normalization in OSPA. We also discuss the choice of GOSPA parameters. 

\begin{defn}\label{def:GOSPA}
Let $c>0$, $0 < \alpha \leq 2$ and $1\leq p < \infty$. Let $d(x, y)$ denote a metric for any $x, y \in \RN$ and let $\dc(x, y) = \min(d(x, y), c)$ be its cut-off metric \cite[Sec.~ 6.2.1]{mahler2014advances}. Let $\Pi_n$ be the set of all permutations of $\{1, \ldots, n\}$ for any $n\in \mathbb{N}$ and any element $\pi \in \Pi_n$ be a sequence $(\pi(1), \ldots, \pi(n))$. Let $X = \{x_1, \ldots, x_{\nX}\}$ and $Y = \{y_1, \ldots, y_{\nY}\}$ be finite subsets of $\RN$. For $\nX \leq \nY$, the GOSPA metric is defined as
\begin{flalign}
&\dG(X, Y) &\nn\\
& \triangleq \left(\min\limits_{\pi \in \Pi_{\nY}} \sum\limits_{i=1}^{\nX} \dc(x_i, y_{\pi(i)})^p + \frac{c^p}{\alpha}(\nY - \nX)\right)^\frac{1}{p}.& \label{eqn:GOSPAdef}
\end{flalign}
If $\nX > \nY$, $\dG(X, Y) \triangleq \dG(Y, X)$.\qed
\end{defn}

It can be seen from the definition that the non-negativity, symmetry and definiteness properties of a metric hold for GOSPA. The proof of the triangle inequality is provided in Appendix~\ref{app:triInProof}.

We briefly discuss the roles of the parameters $p$, $c$ and $\alpha$. The role of the exponent $p$ in GOSPA is similar to that in OSPA \cite{schuhmacher2008consistent}. The larger the value of $p$ is, the more the outliers are penalized. The parameter $c$ in GOSPA determines the maximum allowable localization error and, along with $\alpha$, it also determines the error due to cardinality mismatch. By setting the parameter $\alpha = 1$, we get the OSPA metric without normalization, which divides the metric by $\max\big(|X|,|Y|\big)$. In Section~\ref{sec:issueNorm}, we first discuss why the normalization in OSPA should be removed. In Section~\ref{sec:alpha2}, we indicate the most suitable choice of $\alpha$ for evaluating MTT algorithms..

%\begin{rem}
%The region, where the GOSPA metric is defined, is over finite subsets of entire $\RN$. On the contrary, in OSPA, we require the definition of a bounded window $W \subset \RN$, see \cite[Sec. II]{schuhmacher2008consistent}. This bounded window $W$ is necessary in the triangle inequality proof of the OSPA metric in which dummy points chosen outside $W$ are used \cite[App.]{schuhmacher2008consistent}. In the GOSPA proof, we do not use the dummy points. The GOSPA proof includes the unnormalized OSPA, and it can also be extended to OSPA so that the constraint $W$ is not necessary. \hlc{This proof is also provided as part of the supplementary material.}
%\end{rem}

\subsection{On the removal of normalization}\label{sec:issueNorm}
In this section, we illustrate that the normalization in OSPA provides counterinutitive results using the below example.
%It is clear that one can obtain unnormalized OSPA from GOSPA by setting $\alpha = 1$ in \eqref{eq:GOSPAdef}. In this subsection, we illustrate that the normalization $\frac{1}{|Y|}$ in the OSPA metric can lead to counter-intuitive results. This effect can be illustrated with a simple example. 

\begin{example}\label{ex:1}
Let us say the ground truth is $X =\oslash$ and we have estimates $Y_j=\{y_1, , \ldots, y_{j}\}$ indexed with $j \in \mathbb{N}$. Intuitively, for increasing values of $j$, there is a higher number of false targets, so the distance from $X$ to $Y_{j}$ should also increase. However, the OSPA metric is $c$ for any $j \geq 1$. That is, according to the OSPA metric, all these estimates are equally accurate, which is not the desired evaluation in MTT. \qed
\end{example}

This undesirable property of the OSPA metric is due to the normalization. If we remove this normalization from OSPA, the distance is $j^{\frac{1}{p}} c$, which increases with $j$. This example is a clear motivation as to why the normalization should be removed from the OSPA metric to evaluate MTT algorithms. We refer to the OSPA metric without normalization,  i.e., GOSPA with $\alpha = 1$, as `unnormalized OSPA'. The OSPA metric without the normalization has been used in \cite[Sec.~ IV]{williams2015efficient} to obtain minimum mean OSPA estimate. Even though \cite{williams2015efficient} makes use of the unnormalized OSPA as a cost function, it has not been previously proved that it is a metric.

%\subsection{Discussion}
%The role of the parameters $p$ and $c$ in GOSPA are similar to that in OSPA \cite{schuhmacher2008consistent}. In this subsection, we discuss the role of the parameters $c$ and $\alpha$ and analyze how, similar to OSPA, it is possible to view the overall distance as the sum of localization and cardinality error components.
%
%The connection between the parameters $c$ and $\alpha$ to these costs is easier to analyze when we set $p=1$. In this case, the upper bound on $\alpha$, i.e., $\alpha = 2$ implies that the penalty for cardinality mismatch should at least be $\frac{c}{2}$ for every additional point in the larger set. In other words, the cost for cardinality mismatch should at least be half the maximum allowable localization error. If we make this cost smaller, we no longer have a metric as the triangle inequality is not met. On the other hand, for a localization limit $c$, the cardinality mismatch penalty can be set arbitrarily high by changing $\alpha$. 
%
%We want to highlight two choices of $\alpha$. When $\alpha = 1$, GOSPA reduces to unnormalized OSPA, thus proving that the unnormalized OSPA is a metric. As will be explained in the next section, when $\alpha = 2$, the GOSPA metric can be viewed as the sum of the localization errors of the localized targets plus the error due to missed and false targets. We will also argue that the penalty for cardinality mismatch (or missed/false target), $\frac{c^p}{2}$, with this selection is more suitable for MTT than $\alpha=1$ in which the cost for cardinality mismatch is the same as the cut-off $c^p$.

\subsection{Motivation for setting $\alpha=2$ in MTT}	\label{sec:alpha2}

In this section, we argue that the choice of $\alpha =2$ in GOSPA is the most appropriate one for MTT algorithm evaluations. We show that with this choice, the distance metric can be broken down into localization errors for properly detected targets, which are assigned to target estimates, and the error due to missed and false targets, which are left unassigned as there is no correspondence in the other set. This is in accordance with classical performance evaluation methods for MTT \cite[Sec. 13.6]{blackman1999design},\cite{fridling1991performance,rothrock2000performance, mabbs1993performance,Drummond92}.

For the sake of this discussion, we assume that $X$ is the set of true targets and $Y$ is the estimate, though the metric is of course symmetric. Let us consider $x\in X$ and $y\in Y$, such that all the points in $Y$ are far from $x$ and all the points in $X$ are far from $y$. In this case, the target $x$ has been missed and the estimator has presented a false target $y$. Following \cite{Drummond92}, we refer to these two targets as unassigned targets, even though they may or may not be associated to another target in the permutation in \eqref{eqn:GOSPAdef}. If one of these unassigned targets is not associated to another target in the permutation in \eqref{eqn:GOSPAdef}, it contributes with a cost $c^{p}/\alpha$. On the other hand, if two unassigned targets $x$ and $y$ are associated to each other in the permutation in \eqref{eqn:GOSPAdef}, the cost contribution of the pair is $\dc(x,y)=c^{p}$. 

The basic idea behind selecting $\alpha = 2$ is that the cost for a single unassigned (missed or false) target should be the same whether or not it is associated to another target in the permutation in \eqref{eqn:GOSPAdef}. Therefore, given that a pair of unassigned targets costs $c^p$ and an unassigned target costs $c^p/\alpha$, we argue that $\alpha = 2$ is the most appropriate choice. Due to the importance of choosing $\alpha =2$, from this point on, whenever we write GOSPA, we refer to GOSPA with $\alpha =2$, unless stated otherwise.

In GOSPA, any unassigned (missed or false) target always costs $c^{p}/2$, and, as we will see next, GOSPA contains localization errors for properly detected targets and a cost $c^p/2$ for unassigned targets. In fact, GOSPA can be written in an alternative form, which further highlights the difference with OSPA and clarifies the resemblance with classical MTT evaluation methods. 

To show this, we make the assignment/unassignment of targets explicit by reformulating the GOSPA metric in terms of 2D assignment functions \cite[Sec.~6.5]{blackman1999design} \cite[Chap.~17]{schrijver2003combinatorial} instead of permutations. An assignment set $\gamma$ between the sets $\{1, \ldots, \nX\}$ and $\{1, \ldots, \nY\}$ is a set that has the following properties: $\gamma \subseteq \{1, \ldots, \nX\} \times \{1, \ldots, \nY\}$, $(i,j), (i, j^\prime) \in \gamma$ $\implies$ $j = j^\prime$ and $(i,j), (i^\prime, j)\in \gamma$ $\implies$ $i = i^\prime$, where the last two properties ensure that every $i$ and $j$ gets at most one assignment. Let $\Gamma$ denote the set of all possible assignment sets $\gamma$. Then, we can formulate the following proposition.

\begin{prop}\label{prop:GOSPA_assignments}
The GOSPA metric, for $\alpha = 2$, can be expressed as an optimisation over assignment sets  
\begin{align*}
	&\dGptwo(X, Y) &\nn \\
	&= \biggl[\min\limits_{\gamma \in \Gamma} \biggl(\sum\limits_{(i,j)\in \gamma}d(x_i, y_j)^p + \frac{c^p}{2}\bigl( \nX + \nY - 2 |\gamma|\bigr)\biggr)\biggr]^{\frac{1}{p}}. & 
\end{align*}
\begin{proof}
	See Appendix~\ref{app:prop1}. 
\end{proof}
\end{prop}

This proposition confirms that GOSPA penalizes unassigned targets and localization errors for properly detected targets. The properly detected targets and their estimates are assigned according to the set $\gamma$ so the first term represents their localization errors. Missed and false targets are left unassigned, as done in \cite{Drummond92}, and each of them is penalized by $c^p/2$. To understand this, we first note that $|\gamma|$ is the number of properly detected targets. Hence, $|X|-|\gamma|$ and $|Y|-|\gamma|$ represent the number of missed and false targets, respectively, and the term $c^p (|X|+|Y|-2|\gamma|)/2$ therefore implies that any missed or false target yields a cost $c^p/2$. It should also be noted that the notion of cut-off metric $\dc(\cdot, \cdot)$ is not needed in this representation and there is not a cardinality mismatch term. Also, we remark that this representation cannot be used for OSPA or GOSPA with $\alpha\neq2$. We illustrate the choice of $\alpha=2$ in GOSPA and compare it with OSPA in the following example.  

\begin{figure}[t]
\begin{center}
\fbox{
\begin{minipage}[t][4cm][b]{0.451\linewidth}
\centering
\begin{tikzpicture}[style=thick, x=1cm,y=1cm, every text node part/.style={align=center}, every node/.style={font=\small}, scale = 0.7]
\def \len{1.9}
\node[label = below:$x_1$] (cx1) at(0, 2 +\len){$\times$};
\node[label = below:$x_2$] (cx2) at(1.3, 2+\len){$\times$};
\node[label = above:$y_1$] (cy1) at(0, 3+\len){$\bullet$};
\node[label = above:$y_2$] (cy2) at(1.3, 5+\len){$\bullet$};
\draw[dashed](cx1)--(cy1)node[draw = none, fill = none, midway, left]{$\dc = \Delta$};
\draw[dashed](cx2)--(cy2)node[draw = none, fill = none, midway, right]{$\dc = c$};
\end{tikzpicture}
\subcaption{$Y_a = \{y_1, y_2\}$.}
\label{fig:adjOSPAeg1}
\end{minipage}}
\fbox{
\begin{minipage}[t][4cm][b]{0.451\linewidth}
\centering
\begin{tikzpicture}[style=thick, x=1cm,y=1cm, every text node part/.style={align=center}, every node/.style={font=\small}, scale = 0.7]
\def \len{1.9}
\node[label = below:$x_1$] (cx1) at(0, 2 +\len){$\times$};
\node[label = below:$x_2$] (cx2) at(1.3, 2.5+\len){$\times$};
\node[label = above:$y_1$] (cy1) at(0, 3+\len){$\bullet$};
\draw[dashed](cx1)--(cy1)node[draw = none, fill = none, midway, left]{$\dc = \Delta$};
\end{tikzpicture}
\subcaption{$Y_b = \{y_1\}$.}
\label{fig:adjOSPAeg2}
\end{minipage}}
\caption{$\times$`s are the points in set $X$ and $\bullet$`s are the points in $Y$. The permutation associations are shown using the dashed lines and the cut-off distances are shown on top of them. In this illustration, $\Delta < c$.}
\label{fig:adjOSPAeg}
\end{center}
\end{figure}
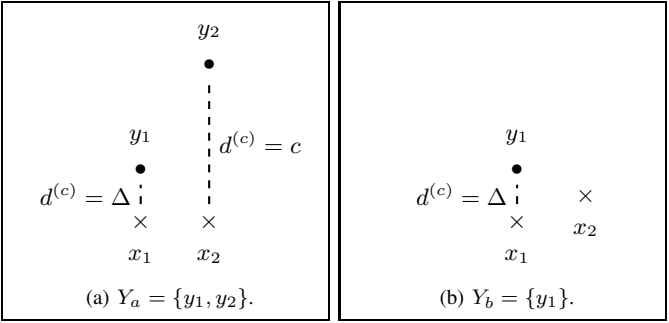

\begin{example}\label{ex:2}
Consider the case where the ground truth is $X = \{x_1, x_2\}$ and there are two estimates $Y_a = \{y_1, y_2\}$ and $Y_b  = \{y_1\}$, as illustrated in Figures~\ref{fig:adjOSPAeg1} and~\ref{fig:adjOSPAeg2}. Targets $x_{2}$ and $y_{2}$ are very far away so that it is obvious that $y_{2}$ is not an estimate of $x_{2}$. Clearly, besides the localization error between $x_1$ and $y_1$, the estimate $Y_a$ has missed target $x_2$ and reported a false target $y_2$, whereas $Y_b$ has only missed target $x_2$. OSPA and unnormalized OSPA provide the same distance to the ground truth for both estimates, $\frac{\Delta + c}{2}$ and $\Delta+ c$, respectively. As a result, according to these metrics, both estimates are equally accurate, which does not agree with intuition and classical MTT evaluation methods. On the contrary, the GOSPA metric shows a desirable trend since $\dGtwo(X, Y_a) = \Delta + c$ is larger than $\dGtwo(X, Y_b) = \Delta + \frac{c}{2}$. \qed
\end{example}

\section{Performance evaluation of MTT algorithms}\label{sec:perf}
In the previous section, we studied metrics between finite sets of targets. It was then implicitly assumed that the ground truth and the estimates are deterministic. However, MTT is often formulated as a Bayesian filtering problem where the ground truth is an RFS and the estimates are sets, which depend deterministically on the observed data \cite{mahler2007statistical}. For performance evaluation, in many cases, we average over several realizations of the data, so estimates are RFSs as well. Therefore, evaluating the performance of several algorithms is in fact a comparison between the RFS of the ground truth and the RFSs of the estimates. As in the case of deterministic sets \cite[pp. 142]{mahler2014advances}, it is highly desirable to establish metrics for RFSs for performance evaluation, which is the objective of this section. We begin with a discussion on the metrics for vectors and random vectors case, and then show how we use these concepts to extend the GOSPA metric to RFSs. 

%In this section, we show that the average GOSPA, $\E[\dG(X, Y)]$ is a metric between RFSs, where the expectation is with respect to the joint PDF of RFSs $X$ and $Y$. In fact, we show that a generalized version, $\sqrt[p^{\prime}]{\E[\dG(X, Y)^{p^{\prime}}]}$, where $1\leq p^{\prime} <\infty$, is a metric. The case when $p^{\prime}=2$ can be viewed as an extension of the root mean square error (RMSE) metric between random vectors \cite[Sec.~ 2.2]{rachev2013methods} to RFSs.

There are several metrics in the literature for random vectors $x,y \in \RN$. If we have a metric in $\RN$, we have a metric on random vectors in $\RN$ by taking the expected value \cite[Sec.~ 2.2]{rachev2013methods}. Then, a natural choice is to compute the average Euclidean distance, $\E[\|x-y\|_2] \triangleq  \int \int \|x-y\|_2 f(x,y)\: dx\: dy$, as a metric on random vectors, where $\|x-y\|_2$ is the Euclidean distance and $f(x,y)$ is the joint density of $x$ and $y$. Another popular metric for random vectors is the root mean square error (RMSE) metric,  $\sqrt{\E[\|x-y\|_2^2]}$ \cite[Sec.~ 2.2]{rachev2013methods}. An advantage with the RMSE, compared to the average Euclidean error, is that it is easier to use it to construct optimal estimators, since it is equivalent to minimizing the mean square error (MSE); note that the MSE, $\E[\|x-y\|_2^2]$, and the squared Euclidean distance $\|x-y\|_2^2$ are not metrics. 

Similar to the Euclidean metric for vectors, one can use the GOSPA metric defined over finite sets to define metrics over RFSs. Following the approaches in the random vector case, root mean square GOSPA and mean GOSPA seem like natural extensions to RFSs. Below, we establish a more general metric for RFS based on GOSPA for arbitrary $\alpha$. 

\begin{prop}\label{prop:RFS}
For $1\leq p, p^{\prime} < \infty$, $c > 0$ and $0 < \alpha \leq 2$, $\sqrt[p^{\prime}]{\E \left[ \dG(X, Y)^{p^{\prime}}\right]}$ is a metric for RFSs $X$ and $Y$.
\begin{proof}
See Appendix~\ref{app:avgGospa}. 
\end{proof}
\end{prop}

For the GOSPA analogue of RMSE, one can set $p^{\prime}= p=2$ and use Euclidean distance for $d(\cdot, \cdot)$. Similar to the minimum MSE estimators in random vectors, one can equivalently use the mean square GOSPA $\E \left[ \dGtwotwo(X, Y)^2\right]$ for obtaining sound RFS estimators based on metrics.

In the RFS case, there are estimators that are obtained by minimizing the mean square OSPA \cite{crouse2011approximate, baum2012calculating, guerriero2010shooting, garcia2011particle} with $p=2$ (or equivalently root mean square OSPA) and Euclidean distance as base metric. One can extend the proof of the proposition to show that the root mean square OSPA is also a metric, which has not been previously established in the literature.

\section{Illustrations}\label{sec:ill}

\begin{figure}[t]
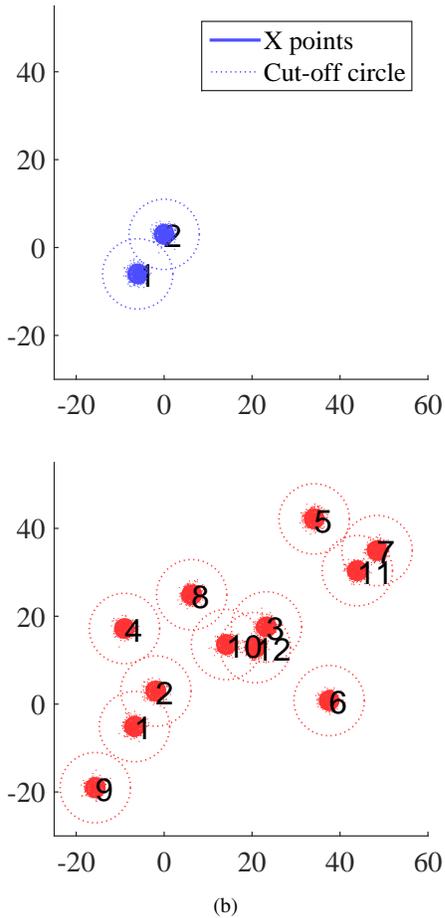

\begin{minipage}[t]{0.9\linewidth}
\centering
\includegraphics[scale=0.6]{truth.eps}
\subcaption{}
\label{fig:truth}
\end{minipage}
\begin{minipage}[t]{0.9\linewidth}
\centering
\includegraphics[scale=0.6]{data_tgt2_fa10.eps} 
\subcaption{}%The red points represent samples from the Bernoulli RFS in $Y$ and the circles represent the data and the 2-sigma Gaussian ellipses. %The dashed circles denote the $c$-radius from the center of these Gaussian components. 
\label{fig:meanGOSPAillus} 
\end{minipage}
\caption{The samples of the ground truth $X$ and estimate $Y$ are illustrated in Figure~\ref{fig:truth} and~\ref{fig:meanGOSPAillus}. The estimate $Y$ has $10$ false targets with indexes from $3$ to $12$ and $2$ properly detected targets with indexes $1$ and $2$ corresponding to the two true targets in Figure~\ref{fig:truth}.}
\end{figure}

\begin{table*}[ht!]
\caption{Table capturing the trends shown by mean metric and root mean square metric for varying number of missed and false targets.}
\centering
\begin{tabular}{|l||c||r|r|r||r|r|r|}
\hline \hline
\multicolumn{2}{|c||}{} & \multicolumn{3}{|c||}{$p^{\prime}=p=1$} & \multicolumn{3}{c|}{$p^{\prime}=p=2$} \\
\hline \hline
\multicolumn{1}{|c||}{} & \shortstack{\scriptsize \# misses $\rightarrow$\\ \scriptsize\# false  $\downarrow$ } & 0 & 1 & 2 & 0 & 1 & 2 \\
\hline \hline

\multirow{4}{*}{\rotatebox[origin = c]{90} {\shortstack{GOSPA \\ $\alpha =2$}}} & 0  & 4.55 & 6.05 & 8 & 3.60 & 6.10 & 8\\
& 1  & 8.62 & 10.04 & 12 & 6.72 & 8.32 & 9.79\\
& 3  & 16.52 & 18.07 & 20 & 10.42 & 11.54 & 12.64 \\
& 10  & 44.49 & 46.05 & 48 &  18.23 & 18.90 & 19.59\\
\hline \hline

\multirow{4}{*}{\rotatebox[origin = c]{90} {OSPA}} & 0  & 2.27 & {\color{blue}5.02} & {\color{blue}8} &  2.55 & {\color{blue}5.88} & {\color{blue}8} \\
& 1  & 4.20 & {\color{blue}5.02} & {\color{blue}8} & 5.07 & {\color{blue}5.88} & {\color{blue}8}\\
& 3  & 5.70 & 6.51 & {\color{blue}8} & 6.39 & 7.02 & {\color{blue}8}\\
& 10 & 7.04 & 7.45 & {\color{blue}8} & 7.37 & 7.65 & {\color{blue}8}\\
\hline \hline

\multirow{4}{*}{\rotatebox[origin = c]{90} {\shortstack{Unnormali- \\ -zed OSPA}}} & 0  & 4.55  & {\color{blue}10.04} & {\color{blue}16} & 3.60 & {\color{blue}8.32} & {\color{blue}11.31}\\
& 1 & 12.62 & {\color{blue}10.04} & {\color{blue}16} & 8.79 & {\color{blue}8.32} & {\color{blue}11.31}\\
& 3 & {\color{red}28.52} & {\color{red}26.07}  & {\color{red} 24} & {\color{red}14.30} & {\color{red}14.04} & {\color{red}13.85}\\ 
& 10  & {\color{red}84.49} & {\color{red}82.05} & {\color{red}80} & {\color{red}25.54} & {\color{red}25.40} & {\color{red}25.29}\\
\hline \hline 

\end{tabular}
\label{tab:res}
\end{table*}

In this section, we show how GOSPA with $\alpha =2$ presents values that agree with the intuition and the guidelines of classical MTT performance evaluation algorithms \cite{rothrock2000performance}, while OSPA and unnormalized OSPA metrics do not. We illustrate these results for  several examples with varying number of missed and false targets in the estimates.

As mentioned in Section ~\ref{sec:perf}, in a Bayesian setting, both the ground truth and estimates are RFSs and we want to determine which estimate is closest to the ground truth. Rather than providing a full MTT simulation, we assume that the ground truth and estimates are specific RFSs, which are easy to visualize and are useful to illustrate the major aspects of the proposed metrics. 

We consider a ground truth $X$ (see Figure~\ref{fig:truth})  which is a multi-Bernoulli RFS  \cite[Sec. 4.3.4]{mahler2014advances} composed of two independent Bernoulli RFSs, each with existence probability $1$. The probability densities of the individual RFSs are Gaussian densities $\mathcal{N}\left([-6,\: -6]^t, \eye \right)$ and $\mathcal{N}\left([0,\:3]^t, \eye \right)$ where $I$ denotes the identity matrix and the notation $v^t$ denotes the transpose of the vector $v$. Therefore, there are always two targets present, which are distributed independently with their corresponding densities. 

We consider scenarios with different estimates $Y$ for this ground truth. By varying $Y$, the number of missed and false targets in each scenario is chosen from $\{0, 1, 2\}$ and $\{0, 1, 3, 10\}$, respectively.  In all the cases, the estimate $Y$ is also a multi-Bernoulli RFS, that contains the Bernoulli sets depicted in Figure~\ref{fig:meanGOSPAillus}. The components with indexes $1$ and $2$ are Gaussian components with densities $\mathcal{N}\left([-6.7, \:-5.1]^{t}, I\right)$ and $\mathcal{N}\left([-1.8,\: 2.9]^{t}, I\right)$ and correspond to estimates of the targets in the ground truth in Figure~\ref{fig:truth}. The remaining components, with indexes $3$ to $12$, are false targets. In scenarios where there is one missed target, we consider that component $1$ has existence probability $1$ but component $2$ has $0$ probability. In scenarios where there are not any missed targets, we consider that both components have existence probability $1$. In the scenarios where the estimate reports $n$ false targets, the existence probability takes the value $1$  for the components $3$ to $n+2$ and  the value $0$ for the remaining false target components. 

We compute the GOSPA and OSPA metrics for RFSs in Proposition~\ref{prop:RFS} for the above scenarios and average the metric values over $1000$ Monte Carlo points. We set $c=8$ and the value of $p^{\prime}=p$ is chosen from $\{1, 2\}$. The Euclidean metric is used as the base distance $d(\cdot, \cdot)$. The estimation errors of these scenarios are tabulated in Table~\ref{tab:res}. The table has estimates with increasing number of missed targets when traversed across columns and increasing number of false targets when traversed across rows. 

Let us first analyze the behavior of the different metrics for varying number of missed targets. Intuitively, as one traverses across columns, the distance between the RFSs should increase with increasing number of missed targets. This trend is observed with GOSPA and the OSPA metric for both $p^{\prime}=p=1$ and $p^{\prime}=p=2$, but the unnormalized OSPA metric shows undesired behaviors when there are false targets in the scenarios (the entries with red text in Table~\ref{tab:res}). To explain this, we look at the expression for the unnormalized OSPA in these scenarios. If $n$ and $m$ are the number of  false and missed targets, and $d_1 < c$ and $d_2 < c$ are the cut-off distances for the properly detected targets, then the unnormalized OSPA when $n\geq 2$ is
\begin{flalign}
 \dGpone = \begin{cases} (n c^p)^\frac{1}{p} & m = 2 \\ 
 (d_1^p + n c^p)^\frac{1}{p} & m = 1 \\
  (d_1^p + d_2^p + n c^p)^\frac{1}{p} & m =0
  \end{cases}.
\end{flalign} 
For $n=1$, $\dGpone$ takes the values $c$, $(d_1^p +c^p)^{\frac{1}{p}}$ and $(d_1^p + d_2^p +c^p)^{\frac{1}{p}}$ respectively. Clearly, for fixed $n\geq 2$, $\dGpone$ decreases with increasing number of missed targets $m$, which is not desirable.

Let us now analyze the behavior of the metrics for varying number of false targets. As the number of false targets increases, the metric should increase \cite{Drummond92}. This trend is displayed by GOSPA for $\alpha =2$ again. On the other hand, the unnormalized OSPA shows a non-decreasing behavior (the entries with blue text in Table~\ref{tab:res}), which is not desirable. The OSPA metric also shows counter intuitive behavior as we discussed in Example~\ref{ex:1} in Section ~\ref{sec:issueNorm}. When both targets are missed, the OSPA metric is constant for varying number of false targets. Also, for the case with one missed target, OSPA and unnormalized OSPA have the same metric values when there are no false targets and when there is one false target. This trend is similar to the behavior we observed in Example~\ref{ex:2}.

\section{Conclusions}\label{sec:concl}
In this paper, we have presented the GOSPA metric. It is a metric for sets of targets that penalizes localization errors for properly detected targets and missed and false targets, in accordance with the classical MTT performance evaluation methods. In difference to the OSPA metric, the GOSPA metric therefore encourages trackers to have as few false and missed targets as possible. 

In addition, we have extended the GOSPA metric to the space of random finite sets of targets. This is important for performance evaluation of MTT algorithms.

\appendices
\input{supplementary31Jan2017.tex}

\bibliographystyle{IEEEtranS}
\input{GOSPA_Fusion31Jan2017.bbl}

\end{document}

%% file: supplementary31Jan2017.tex
\section{Proof of the triangle inequality of GOSPA}\label{app:triInProof}
In the proof, an extension of Minkowski's inequality \cite[pp. 165]{kubrusly2011elements} to sequences of different lengths by appending zeros to the shorter sequence is used. Let us say we have two sequences $(a_i)_{i=1}^{m}$ and $(b_i)_{i=1}^{n}$ such that $m\leq n$. We extend the sequence $(a_i)$ such that $a_i = 0$ for $i = m+1, \ldots, n$. Then, using Minkowski's inequality on this extended sequence we get that
\begin{flalign}
\left(\sum\limits_{i=1}^{m}|a_i + b_i|^p + \sum\limits_{i=m+1}^{n}|b_i|^p \right)^{\frac{1}{p}} \nn\\
\leq \left( \sum\limits_{i=1}^{m}|a_i|^p\right)^{\frac{1}{p}} +\left(\sum\limits_{i=1}^{n}|b_i|^p \right)^{\frac{1}{p}} \label{eqn:minFinSum}
\end{flalign}
for $1 \leq p < \infty$. We use this result several times in our proof. 

We would like to prove the triangle inequality:
\begin{flalign}
\dG(X, Y) \leq \dG(X, Z) + \dG(Y, Z)
\end{flalign}
for any three RFSs $X$, $Y$ and $Z$. The proof is dealt in three cases based on the values of $\nX$, $\nY$ and $\nZ$. Without loss of generality, we assume $\nY \geq \nX$ in all the three cases, since GOSPA is symmetric in $X$ and $Y$.

\subsection*{Case 1: $\nX \leq \nY \leq \nZ$}
For any $\pi \in \Pi_{\nY}$, 
\begin{flalign}
&\dG(X,Y) \leq \left(\sum\limits_{i=1}^{\nX} \dc(x_i, y_{\pi(i)})^p + \frac{c^p}{\alpha}(\nY - \nX) \right)^{\frac{1}{p}}.&
\end{flalign}
Using the triangle inequality on the cut-off metric $\dc(\cdot, \cdot)$, we get that for any $\pi \in \Pi_{\nY}$ and for any $\sigma \in \Pi_{\nZ}$, 
\begin{flalign}
&\dG(X,Y)\leq  \left(\sum\limits_{i=1}^{\nX} \left[\dc(x_i, z_{\sigma(i)}) + \dc(z_{\sigma(i)}, y_{\pi(i)})\right]^p \right.& \nn\\
& \qquad\left. + \frac{c^p}{\alpha}(\nY - \nX) \right)^{\frac{1}{p}}  &\\
&\leq  \left(\sum\limits_{i=1}^{\nX} \left[\dc(x_i, z_{\sigma(i)}) + \dc(z_{\sigma(i)}, y_{\pi(i)})\right]^p  \nn\right. &\\ 
& \qquad+ \frac{c^p}{\alpha}(\nY - \nX) + 2 \frac{c^p}{\alpha}(\nZ - \nY) &\nn\\
&\qquad\left. +\sum\limits_{i=\nX+ 1}^{\nY} \dc(z_{\sigma(i)}, y_{\pi(i)})^p  \right)^{\frac{1}{p}}  & \label{eqn:Gospaproofi1}\\
&=  \left(\sum\limits_{i=1}^{\nX} \left[\dc(x_i, z_{\sigma(i)}) + \dc(z_{\sigma(i)}, y_{\pi(i)})\right]^p \nn\right. &\\ 
&\qquad\left. + \frac{c^p}{\alpha}(\nZ - \nX)+\sum\limits_{i=\nX+ 1}^{\nY} \dc(z_{\sigma(i)}, y_{\pi(i)})^p \right. &\nn \\
& \qquad\left. + \frac{c^p}{\alpha}(\nZ - \nY) \right)^{\frac{1}{p}}  &\\
& \leq  \left(\sum\limits_{i=1}^{\nX} \dc(x_i, z_{\sigma(i)})^p + \frac{c^p}{\alpha} (\nZ- \nX) \right)^{\frac{1}{p}} &\nn\\
& \qquad+ \left(\sum\limits_{i=1}^{\nY} \dc(z_{\sigma(i)}, y_{\pi(i)})^p  +  \frac{c^p}{\alpha}(\nZ - \nY)\right)^{\frac{1}{p}}.  &
 \end{flalign}
To arrive at the last inequality, Minkowski's inequality in \eqref{eqn:minFinSum} is used. Since $\pi$ is a bijection, we can invert $\pi$ to arrive at 
 \begin{flalign} 
&\dG(X,Y) \leq \left(\sum\limits_{i=1}^{\nX} \dc(x_i, z_{\sigma(i)})^p + \frac{c^p}{\alpha} (\nZ- \nX) \right)^{\frac{1}{p}} &\nn\\
&\quad + \left(\sum\limits_{i=1}^{\nY} \dc(z_{\pi^{-1}(\sigma(i))}, y_{i})^p  +  \frac{c^p}{\alpha}(\nZ - \nY)\right)^{\frac{1}{p}}.&
\end{flalign}
The composition $\pi^{-1}\circ \sigma$ will be a permutation on $\{1, \ldots, \nZ\}$. Lets denote this as $\tau$. So, for any $\tau, \sigma \in \Pi_{\nZ}$,
\begin{flalign}
& \dG(X,Y) \leq \left(\sum\limits_{i=1}^{\nX} \dc(x_i, z_{\sigma(i)})^p + \frac{c^p}{\alpha} (\nZ- \nX) \right)^{\frac{1}{p}} &\nn\\
  &\quad+ \left(\sum\limits_{i=1}^{\nY} \dc(z_{\tau(i)}, y_{i})^p  +  \frac{c^p}{\alpha}(\nZ - \nY)\right)^{\frac{1}{p}},&
\end{flalign}
which also holds for the $\sigma$ and $\tau$ that minimizes the first and the second term in the right hand side. This proves the triangle inequality for this case.

\subsection*{Case 2: $\nX \leq \nZ \leq \nY$}
As before, for any $\pi \in \Pi_{\nY}$ and $\sigma \in \Pi_{\nZ}$,
\begin{flalign}
&\dG(X, Y)\leq   \left(\sum\limits_{i=1}^{\nX} \left[\dc(x_i, z_{\sigma(i)}) + \dc(z_{\sigma(i)}, y_{\pi(i)})\right]^p \right. &\nn\\
& \qquad \left.  + \frac{c^p}{\alpha}(\nY - \nX) \right)^{\frac{1}{p}}&\\
& \leq   \left(\sum\limits_{i=1}^{\nX} \left[\dc(x_i, z_{\sigma(i)}) + \dc(z_{\sigma(i)}, y_{\pi(i)})\right]^p \right. &\nn \\
&\qquad\left.  +\sum\limits_{i=\nX+1}^{\nZ} \dc(z_{\sigma(i)}, y_{\pi(i)})^p  +\frac{c^p}{\alpha}(\nZ - \nX)  \right. &\nn\\
&\qquad\left. +\frac{c^p}{\alpha}(\nY - \nZ) \right)^{\frac{1}{p}} &\\
&\leq  \left(\sum\limits_{i=1}^{\nX} \dc(x_i, z_{\sigma(i)})^p  +\frac{c^p}{\alpha}  (\nZ - \nX) \right)^{\frac{1}{p}}  &\nn\\
&\qquad+\left(\sum\limits_{i=1}^{\nZ} \dc(z_{\sigma(i)}, y_{\pi(i)})^p+ \frac{c^p}{\alpha} (\nY - \nZ) \right)^{\frac{1}{p}}.&
\end{flalign}
From here, we can argue similar to the Case 1 and show that the triangle inequality holds.

\subsection*{Case 3: $\nZ \leq \nX\leq \nY $}
\begin{flalign}
&\dG(X,Y)  \leq\left(  \sum\limits_{i=1}^{\nX} \dc(x_i, y_{\pi(i)})^p + \frac{c^p}{\alpha}(\nY - \nX) \right)^{\frac{1}{p}}& \nn \\
&\leq \Bigg(  \sum\limits_{i=1}^{\nZ} \dc(x_i, y_{\pi(i)})^p +\frac{2 c^p}{\alpha}(\nX - \nZ) \nn &\\
& \qquad + \frac{c^p}{\alpha}(\nY - \nX) \Bigg)^{\frac{1}{p}}.
\end{flalign}
To get the above inequality, for $i = \nZ+1, \ldots, \nX$, we used the fact that $\dc(x_i, y_{\pi(i)})^p \leq c^p \leq 2 \frac{c^p}{\alpha}$ when $0 \leq \alpha \leq 2$.
\begin{flalign}
&\dG(X,Y) \leq \left(  \sum\limits_{i=1}^{\nZ} \left[\dc(x_i, z_{\sigma(i)}) + \dc(z_{\sigma(i)}, y_{\pi(i)})\right]^p \right.& \nn \\
& \qquad\left. +\frac{c^p}{\alpha}(\nY - \nZ) + \frac{c^p}{\alpha}(\nY - \nX) \right)^{\frac{1}{p}}.& 
\end{flalign}
From here, the arguments are similar to the ones in the last two cases. \qed

\section{Proof of Proposition 1}\label{app:prop1}

We proceed to prove Proposition 1. Given $X$ and $Y$, each possible
permutation $\pi\in\Pi_{\left|Y\right|}$ in (1) has a corresponding
assignment set $\gamma_{\pi}=\left\{ \left(i,j\right):\,j=\pi\left(i\right)\,\mathrm{and}\,d\left(x_{i},y_{j}\right)<c\right\} $
such that we can write 
\begin{flalign}
 & d_{p}^{\left(c,2\right)}\left(X,Y\right)=\left(\mathrm{min}_{\pi\in\Pi_{\left|Y\right|}}\sum_{\left(i,j\right)\in\gamma_{\pi}}d\left(x_{i},y_{j}\right)^{p}\right.&\nonumber \\
 & \qquad\left.+c^{p}\left(\left|X\right|-\left|\gamma{}_{\pi}\right|\right)+\frac{c^{p}}{2}\left(\left|Y\right|-\left|X\right|\right)\right)^{1/p}&
\end{flalign}
where we have written $d\left(x_{i},y_{j}\right)^{p}$ instead of
$d^{\left(c\right)}\left(x_{i},y_{\pi\left(i\right)}\right)^{p}$
as the distance between the assigned points in $\gamma_{\pi}$ is
smaller than $c$. Also, $\left|X\right|-\left|\gamma{}_{\pi}\right|$
is the number of pairs $\left(x_{i},y_{\pi\left(i\right)}\right)$
for which $d^{\left(c\right)}\left(\cdot,\cdot\right)=c$, and the
second term compensates for the fact that these pairs are not accounted
for when we sum over $(i,j)\in\gamma_{\pi}$. Rearranging terms we
obtain 
\begin{align*}
d_{p}^{\left(c,2\right)}\left(X,Y\right) & =\left(\mathrm{min}_{\pi\in\Pi_{\left|Y\right|}}\sum_{\left(i,j\right)\in\gamma_{\pi}}d\left(x_{i},y_{j}\right)^{p}\right.\\
 & \quad\left.+\frac{c^{p}}{2}\left(\left|Y\right|+\left|X\right|-2\left|\gamma{}_{\pi}\right|\right)\right)^{1/p}.
\end{align*}
As the space of assignment sets $\Gamma$ is bigger than the set of
assignment sets induced by permutations $\pi\in\Pi_{\left|Y\right|}$,
we have  
\begin{flalign}
 & d_{p}^{\left(c,2\right)}\left(X,Y\right)&\nonumber \\
 & \geq\left(\mathrm{min}_{\gamma\in\Gamma}\sum_{\left(i,j\right)\in\gamma}d\left(x_{i},y_{j}\right)^{p}+\frac{c^{p}}{2}\left(\left|Y\right|+\left|X\right|-2\left|\gamma\right|\right)\right)^{1/p}.&\label{eq:inequality1}
\end{flalign}
We have not yet finished the proof as we have obtained an inequality.
Let us consider $\gamma^{\star}$ to be the value of the assignment
set that minimises the distance in Proposition 1. First, 
\begin{flalign}
 & \left(\sum_{\left(i,j\right)\in\gamma^{\star}}d\left(x_{i},y_{j}\right)^{p}+\frac{c^{p}}{2}\left(\left|Y\right|+\left|X\right|-2\left|\gamma^{\star}\right|\right)\right)^{1/p}&\nn\\
 & =\left(\sum_{\left(i,j\right)\in\gamma^{\star}}d^{\left(c\right)}\left(x_{i},y_{j}\right)^{p}+\frac{c^{p}}{2}\left(\left|Y\right|+\left|X\right|-2\left|\gamma^{\star}\right|\right)\right)^{1/p}&
\end{flalign}
due to the fact that otherwise we could construct a better assignment set $\tilde \gamma = \gamma^{\star} \setminus \gamma^c$ where $\gamma^c=\left\{ (i,j) \in \gamma^{\star} : d(x_i,y_j)>c \right\}$. That is, we know that $\gamma^{\star}$ does not contain pairs $\left(i,j\right)$ for which $d\left(\cdot,\cdot\right)>c$.
On the other hand, if two pairs are unassigned in the optimal assignment,
their distance must be $d\left(\cdot,\cdot\right)>c$ so $d^{\left(c\right)}\left(\cdot,\cdot\right)=c$,
as, otherwise, there would be an assignment that returns a lower value
than the optimal one by assigning them. 

We can now construct a corresponding permutation $\pi_{\gamma^{\star}}\in\Pi_{\left|Y\right|}$
as follows: $\pi_{\gamma^{\star}}\left(i\right)=j$ if $\left(i,j\right)\in\gamma^{\star}$
and the rest of the components of $\pi_{\gamma^{\star}}$ can be filled
out arbitrarily as any selection does not change the value of the
previous equation. Then, 
\begin{align*}
 & \left(\sum_{\left(i,j\right)\in\gamma^{\star}}d^{\left(c\right)}\left(x_{i},y_{j}\right)^{p}+\frac{c^{p}}{2}\left(\left|Y\right|+\left|X\right|-2\left|\gamma^{\star}\right|\right)\right)^{1/p}\\
 & =\left(\sum_{i=1}^{\left|X\right|}d^{\left(c\right)}\left(x_{i},y_{\pi_{\gamma^{\star}}\left(i\right)}\right)^{p}+\frac{c^{p}}{2}\left(\left|Y\right|-\left|X\right|\right)\right)^{1/p}\\
 & \leq d_{p}^{\left(c,2\right)}\left(X,Y\right).
\end{align*}
Therefore, we now have proved that 
\begin{flalign}
 & \left(\mathrm{min}_{\gamma\in\Gamma}\sum_{\left(i,j\right)\in\gamma}d\left(x_{i},y_{j}\right)^{p}+\frac{c^{p}}{2}\left(\left|Y\right|+\left|X\right|-2\left|\gamma\right|\right)\right)^{1/p}&\nn\\
 & \leq d_{p}^{\left(c,2\right)}\left(X,Y\right)&
\end{flalign}
which together with (\ref{eq:inequality1}) proves Proposition 1. \qed

\section{Proof of the average GOSPA metric}\label{app:avgGospa}
For RFS $X$ with the multi object density function $f(\cdot)$ and for a real valued function of RFS $g(\cdot)$, using the set integral, the expectation of $g(X)$ \cite[pp. 177]{goodman2013mathematics} is:
\begin{flalign}
&\E [g(X)] =  \int f(X) g(X) \delta X &\\
&=  \sum\limits_{n=0}^{\infty} \frac{1}{n!}\int f(\{x_1, \ldots, x_n\}) g(\{x_1, \ldots,x_n\})d(x_1, \ldots ,x_n) .\nn&
\end{flalign}

Similarly, we have that for random finite sets $X$, $Y$ with joint density $f(\cdot, \cdot)$
\begin{flalign}
&\sqrt[p^{\prime}]{\E[\dG(X, Y)^{p^{\prime}}]}& \nn\\
&\quad =\sqrt[{p^{\prime}}]{\int \int \dG(X, Y)^{p^{\prime}} f(X, Y) \delta X \delta Y}.&\label{eqn:Egospa}
\end{flalign}
Since $X$ and $Y$ are random finite sets, $f(X, Y)$ is non-zero only when $X$ and $Y$ are finite, and in this case $\dG(X, Y)$ is finite. These conditions imply that $\E[\dG(X, Y)^{p^{\prime}}] < \infty$ is satisfied. Definiteness, non-negativity and symmetry properties of \eqref{eqn:Egospa} are observed directly from the definition.  Note that, for metrics in the probability space, the definiteness between random variables is in the almost sure sense \cite[Sec.~ 2.2]{rachev2013methods}. The proof of the triangle inequality is sketched below.

In the proof, we use Minkowski's inequality for  infinite sums and for integrals  \cite[pp. 165]{kubrusly2011elements}. Using these Minkowski's inequalities, we can show that the inequality also extends to cases that have both infinite sums and integrals as it appears in the set integrals. For real valued functions $\psi_n(x_{1:n})$ and $\phi_n(x_{1:n})$ such that $\int \left| \psi_n(x_{1:n}) \right|^{{p^{\prime}}}dx_{1:n} < \infty$  and $\int \left| \phi_n(x_{1:n}) \right|^{{p^{\prime}}}dx_{1:n} < \infty$  for  $n=1, \ldots, \infty$ and ${p^{\prime}}\geq 1$,
\begin{flalign}
&\left(\sum_{n=0}^{\infty}\int\left| \psi_n(x_{1:n})+\phi_n(x_{1:n})\right|^{{p^{\prime}}}dx_{1:n}\right)^{\frac{1}{{p^{\prime}}}}&	\nn \\
\leq &	\left(\sum_{n=0}^{\infty}\int \left| \psi_n(x_{1:n}) \right|^{{p^{\prime}}}dx_{1:n} \right)^{\frac{1}{{p^{\prime}}}}&\nn\\
& +\left(\sum_{n=0}^{\infty}\int \left| \phi_n(x_{1:n})\right|^{{p^{\prime}}}dx_{1:n} \right)^{\frac{1}{{p^{\prime}}}}. &\label{eqn:minSumInt}
\end{flalign}
The inequality in \eqref{eqn:minSumInt} can be proved by first using Minkowski's inequality for integrals on the LHS: 
\begin{flalign}
&\int\left| \psi_n(x_{1:n})+\phi_n(x_{1:n})\right|^{{p^{\prime}}}dx_{1:n}&\nn \\
\leq &\left( \left( \int |\psi_n(x_{1:n})|^{p^{\prime}} dx_{1:n}\right)^{\frac{1}{{p^{\prime}}}} \right. &\nn\\
& \left. + \left( \int |\phi_n(x_{1:n})|^{p^{\prime}} dx_{1:n}\right)^{\frac{1}{{p^{\prime}}}}\right)^{p^{\prime}}.&
\end{flalign}
And then using Minkowski's inequality for infinite sums on this, we get the RHS of \eqref{eqn:minSumInt}.

Now, we use the above results for the triangle inequality of \eqref{eqn:Egospa}. Let us consider RFS $X$, $Y$ and $Z$ with joint distribution $f(X, Y, Z)$:
\begin{flalign}
&\sqrt[{p^{\prime}}]{\E[\dG(X, Y)^{p^{\prime}}]}&\nn\\
&\leq \sqrt[{p^{\prime}}]{\E\left[\left(\dG(X, Z)+ \dG(Z, Y)\right)^{p^{\prime}}\right]} &\\
&=\biggl[ \int \int \int \left(\dG(X, Z)+ \dG(Z, Y)\right)^{p^{\prime}} &\nn\\
& \qquad \times f(X, Y, Z)\delta X \delta Y \delta Z \biggr]^{\frac{1}{{p^{\prime}}}}&\\
&=\left[ \int \int \int \left(\dG(X, Z) f(X, Y, Z)^\frac{1}{{p^{\prime}}} \right.\right. &\nn\\
& \qquad \left. \left. + \dG(Z, Y) f(X, Y, Z)^\frac{1}{{p^{\prime}}}\right)^{p^{\prime}} \delta X \delta Y \delta Z \right]^{\frac{1}{{p^{\prime}}}}.&
\end{flalign}
If we expand the set integrals, they are of the form 
\begin{flalign}
&\sqrt[{p^{\prime}}]{\E[\dG(X, Y)^{p^{\prime}}]} \leq \biggl[ \sum\limits_{i=0}^{\infty} \sum\limits_{j=0}^{\infty} \sum\limits_{k=0}^{\infty}\int \int \int  &\nn\\
& \qquad \bigg(f_1(\{x_1, \ldots, x_i\}, \{y_1, \ldots, y_j\}, \{z_1, \ldots, z_k\}) &\nn\\
& \qquad  + f_2(\{x_1, \ldots, x_i\}, \{y_1, \ldots, y_j\}, \{z_1, \ldots, z_k\})\bigg)^{p^{\prime}}  & \nn \\
& \qquad \times d(x_1, \ldots, x_i) \:d(y_1, \ldots, y_j)\: d(z_1, \ldots, z_k) \biggr]^{\frac{1}{{p^{\prime}}}},& \label{eqn:mgospa_inter1}
\end{flalign}
where 
\begin{flalign}
&f_1(\{x_1, \ldots, x_i\}, \{y_1, \ldots, y_j\}, \{z_1, \ldots, z_k\}) &\nn\\
&= \dG(\{x_1,\ldots, x_i\}, \{z_1, \ldots, z_k\})&  \nn  \\
& \qquad \times \left(\frac{f(\{x_1, \ldots, x_i\}, \{y_1, \ldots, y_j\}, \{z_1, \ldots, z_k\})}{i!\: j! \: k!}\right)^{\frac{1}{{p^{\prime}}}} &
\end{flalign}
and 
\begin{flalign}
&f_2(\{x_1, \ldots, x_i\}, \{y_1, \ldots, y_j\}, \{z_1, \ldots, z_k\}) &\nn\\
&= \dG(\{y_1,\ldots, y_j\}, \{z_1, \ldots, z_k\})&  \nn  \\
& \qquad \times \left(\frac{f(\{x_1, \ldots, x_i\}, \{y_1, \ldots, y_j\}, \{z_1, \ldots, z_k\})}{i!\: j! \: k!}\right)^{\frac{1}{{p^{\prime}}}}. &
\end{flalign}
The multiple integrals and sums in \eqref{eqn:mgospa_inter1} can be considered as one major integral and sum. Using Minkowski's inequality in \eqref{eqn:minSumInt} for infinite sums and integrals  on \eqref{eqn:mgospa_inter1}, we get
\begin{align}
&\sqrt[{p^{\prime}}]{\E[\dG(X, Y)^{p^{\prime}}]}&\nn\\
&\leq\left[ \int \int \int \dG(X, Z)^{p^{\prime}}  f(X, Y, Z)\delta X \delta Y \delta Z \right]^{\frac{1}{{p^{\prime}}}}&\nn\\
&\qquad +\left[  \int \int \int \dG(Y, Z)^{p^{\prime}}  f(X, Y, Z)\delta X \delta Y \delta Z \right]^{\frac{1}{{p^{\prime}}}}&\\
&=\left[ \int \int \dG(X, Z)^{p^{\prime}}  f(X, Z)\delta X \delta Z \right]^{\frac{1}{{p^{\prime}}}}&\nn\\
&\qquad +\left[ \int \int \dG(Y, Z)^{p^{\prime}}  f(Y, Z)\delta Y \delta Z \right]^{\frac{1}{{p^{\prime}}}},&
\end{align}
which finishes the proof of the triangle inequality. \qed